\documentclass[aps,prl,twocolumn,showpacs,groupaddress]{revtex4}
\usepackage{amssymb}
\usepackage{graphicx}
\usepackage{dcolumn}
\usepackage{bm}
\usepackage{amsmath}

\begin{document}

\title{Transform-limited pulses are not optimal for resonant multiphoton transitions }
\author{Nirit Dudovich}
\email[]{Nirit.Dudovich@Weizmann.ac.il}
\author{Barak Dayan}
\email[]{Barak.Dayan@Weizmann.ac.il} \author{S.M.G. Faeder}
\author{Yaron Silberberg}
\homepage[]{www.weizmann.ac.il/home/feyaron/}
\affiliation{Department of Physics of Complex Systems, Weizmann Institute of Science,\\
Rehovot 76100, Israel}

\begin{abstract}
Maximizing nonlinear light-matter interactions is a primary motive
for compressing laser pulses to achieve ultrashort transform
limited pulses. Here we show how, by appropriately shaping the
pulses, resonant multiphoton transitions can be enhanced
significantly beyond the level achieved by maximizing the pulse's
peak intensity. We demonstrate the counterintuitive nature of this
effect with an experiment in a resonant two-photon absorption, in
which, by selectively removing certain spectral bands, the peak
intensity of the pulse is reduced by a factor of 40, yet the
absorption rate is doubled. Furthermore, by suitably designing the
spectral phase of the pulse, we increase the absorption rate by a
factor of 7.
\end{abstract}


\maketitle

The ability to steer quantum systems by coherently manipulating the interacting light
 (coherent quantum control), has been proposed \cite{1,2,3,4} and
 recently demonstrated \cite{5,6,7,8,9,10}
 for manipulating both simple and complex systems such as atoms, molecules and semiconductors.
 Weiner et al. \cite{11} demonstrated selectivity among Raman transitions excited by shaped pulses.
 They showed that by splitting an initially transform limited pulse into a pulse sequence
 with a specific repetition rate, a certain vibrational mode of a molecular crystal was excited.
 Meshulach and Silberberg \cite{12,13} demonstrated that by manipulating the spectral phase of the
 exciting laser pulse, one could reduce and even annihilate the two photon absorption (TPA) rate.
 Nevertheless, it has been established \cite{13} that these processes are maximized by
 transform-limited pulses, and cannot be enhanced beyond that level by shaping the pulses.
 Here we show that for transitions that involve an intermediate resonant state, this limit is no
 longer valid. By shaping the pulses in a way that exploits the spectral response of the interaction
 around the resonance we enhanced the resonant TPA rate in Rb vapor by a factor of 7, even though their
 peak intensity was reduced. In another experiment we enhanced the TPA rate by
 $100\%$ simply by blocking all the red-detuned frequencies of the pulse, although the peak intensity
 of the pulse was reduced by a factor of 40.\\

Consider a TPA in an atomic system, induced by a weak femtosecond
laser pulse with an electric field $\varepsilon(t)$. The amplitude
of the excited state is predicted by the second order time
dependent perturbation theory:

\begin{eqnarray}
\label{eq1}\
a_{f}(t)=&-&\frac{1}{\hbar^{2}}\sum_{n}\mu_{fn}\mu_{ng}\int_{-\infty}^{t}\int_{-\infty}^{t_1}
\varepsilon(t_1)\varepsilon(t_2)\nonumber\\&\times&exp(i\omega_{fn}t_1)\exp(i\omega_{ng}t_2)dt_2dt_1\:,
\end{eqnarray}

where $\mu_{fn}$ and $\mu_{ng}$ are the dipole moment matrix
elements, with $|g\rangle$,$|n\rangle$ and $|f\rangle$ the ground,
intermediate, and final levels,
$\omega_{ij}\equiv(E_i-E_j)/\hbar$, and the summation is performed
over all possible intermediate states of the unperturbed atom. The
pulse duration is assumed here to be considerably shorter than all
lifetimes involved. In a nonresonant TPA, all the intermediate
levels of the atom are considerably far from the pulse frequency
spectrum. The nonresonant excited state amplitude can then be
approximated by \cite{13}:

\begin{eqnarray}
\label{eq2} a_{f}^{nr}\approx&-&\frac{1}{i\hbar^{2}}
\sum_{n}\frac{\mu_{fn}\mu_{ng}}{\omega_{ng}-\omega_{fg}/2}\nonumber\\&\times&\int_{-\infty}^{\infty}
E(\omega)E(\omega_{fg}-\omega)d\omega\:.
\end{eqnarray}

where $E(\omega)$ is the Fourier transform of $\varepsilon(t)$,
and the pulse spectrum is taken to be centered on $\omega_{fg}/2$.
Equation (\ref{eq2}) reflects the fact that two-photon transitions
occur for all pairs of photons with frequencies that additively
give the final transition energy. As is evident from Eq.
(\ref{eq2}), the nonresonant TPA rate is maximized by a transform
limited pulse, where all the spectral elements of $E(\omega)$ have
the same phase, and therefore add constructively.\

In the case of a resonant TPA, some intermediate (resonant) levels
are within the spectral range of the pulse. The contribution to
the excited state amplitude due to a transition through an
intermediate resonant level $|i\rangle$ is given by:

\begin{eqnarray}
\label{eq3}
a_{f}^{r}\approx&-&\frac{1}{i\hbar^{2}}\mu_{fi}\mu_{ig}\bigg[\:i\pi
E(\omega_{ig})E(\omega_{fg}-\omega_{ig})\nonumber\\&+&\wp\int_{-\infty}^{\infty}
\frac{E(\omega)E(\omega_{fg}-\omega)}{\omega_{ig}-\omega} d\omega
\:\bigg]\:.
\end{eqnarray}

where $\wp$ is the principle value of Couchy, and
\mbox{$\omega_{ig},\:\omega_{fg}-\omega_{ig}=\omega_{fi}$} are the
resonance frequencies. As is evident from Eq. (\ref{eq3}), the
resonant process exhibits a different spectral behavior from that
of the nonresonant one. In cases where a single resonant level
exists close to the two-photon frequency $\omega_{fg}/2$ (compared
to the spectral bandwidth of the pulse), the contribution of the
nonresonant process is negligible, therefore Eq. (\ref{eq3})
describes the total TPA rate.\

The first term in Eq. (\ref{eq3}) depends only on the spectral
components of the pulse at the resonance frequencies, whereas the
second term integrates over the contributions of all other
spectral components of the pulse. The broad spectral dependence of
the second term originates from the short time ($\sim10^{-13}$
sec) the atom spends at the intermediate level before absorbing
the second photon, which according to the uncertainty principle
allows for some detuning between the exciting photons and the
resonance frequencies. Nevertheless, the larger the detuning, the
lower the probability the atom will stay at the intermediate level
long enough to absorb a second photon, hence the
$\frac{1}{\omega_{ig}-\omega}$ factor. Due to the considerably
longer time the atom remains at the final level, the frequencies
of all the photon pairs must sum to $\omega_{fg}$, and hence the
dependence on $E(\omega)E(\omega_{fg}-\omega)$ in Eqs. (\ref{eq2})
and (\ref{eq3}). As expected for a harmonically driven system, the
first term (the on-resonance contribution) is shifted by $\pi/2$
compared with the second term(the off-resonance contributions).
Also, the spectral components below and above the resonance excite
the system in-phase and $\pi$ out-of-phase, respectively. We shall
utilize these phase relations to enhance the nonlinear response.\

Using the above derivation, we can predict the TPA rate for
various pulses with different spectra. When the atom is subjected
to a transform limited pulse, the second term in Eq. (\ref{eq3})
integrates over both negative and positive contributions, and
becomes negligible as the spectral width of the exciting pulse
grows larger than $|\omega_{ig}-\omega_{fg}/2|$. In other words, a
transform limited pulse induces a destructive quantum interference
between the events in which the first exciting photon is red
detuned and those in which it is blue detuned. A simple, although
counterintuitive, way to enhance the transition probability is to
prevent this destructive interference by blocking all red (or
blue) detuned photons, despite the fact that this will reduce the
pulse's peak intensity (due to both attenuation of the power and
broadening of the pulse). Larger enhancement can be achieved by
applying a phase function that inverts the sign of
$E(\omega)E(\omega_{fg}-\omega)$ about the resonance, so that all
photon pairs interfere constructively, and the transition
probability is maximized.  Since the integrand in Eq. (\ref{eq3})
approaches its maximum absolute value around the singular point at
$\omega_{ig}$, this enhancement depends on the actual spectral
resolution of the phase function, which limits the sharpness of
the sign inversion around $\omega_{ig}$.\\

To demonstrate these enhancements experimentally, we considered
the resonant TPA in Rubidium gas between the 5S and the 5D states,
which is dominated by resonant transitions through the 5P level
(Fig. 1a). The TPA was induced by pulses with a bandwidth of 18 nm
(corresponding to 50 fs transform limited pulses), produced by a
mode-locked Ti:sapphire laser with an average power of 150 mW and
repetition rate of 80 MHz. The spectrum, centered
on the two-photon transition frequency $\omega_{fg}/2$ (778 nm),
overlapped with the 5S-5P and the 5P-5D resonant
transitions at $\omega_{ig}$ and $\omega_{fi}$ corresponding to
780 nm and 776 nm, respectively. As excited atoms decay
spontaneously to the ground level through the 6P level, the TPA
intensity is evaluated by measuring the fluorescence at ~420 nm.
We adjusted the spectral phase of the pulse with a programmable
pulse shaper, which includes a liquid crystal spatial light
modulator (SLM) at its Fourier plane \cite{14,15,16}. The shaper
enables both cancellation of dispersion, as well as the
application of any desired additional spectral phase mask by
applying different phase shifts to the spatially separated
spectral components of the pulse. Our setup (Fig. 1b) was almost
identical to the one in \cite{12}, except that the input laser
beam was expanded in order to improve the spectral resolution to
less than 0.3 nm.\

\begin{figure}[tbp] \label{fig1}
\begin{center}
\includegraphics[width=8.6cm]{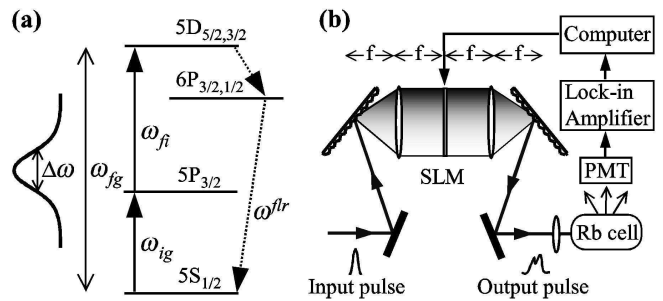}
\end{center}
\caption{(a) Energy levels diagram of a resonant TPA in Rb. The
frequencies of the 5S-5P ($\omega_{ig}$) and 5P-5D ($\omega_{fi}$)
resonant transitions correspond to 780.2 nm and 776.0 nm,
respectively. The pulse spectrum is centered on the two-photon
transition frequency ($\omega_{fg}/2$) at 778.1 nm, with a
bandwidth of $\triangle\omega=$18 nm (FWHM). The excited atoms
spontaneously decay to the ground level through the 6P, emitting a
fluorescence signal at $\omega^{flr}$ ($\approx 420$ nm). (b) The
experimental setup. Femtosecond laser pulses were modified in a
computer-controlled 4-f pulse shaper. The pulse shaper is composed
of a pair of diffraction gratings and a pair of achromat lenses. A
programmable SLM with 128 computer controlled discrete elements is
placed at the Fourier plane and is used to apply phase masks to
the spectrum of the pulse. The shaped pulses were focused into the
Rb cell, and the fluorescence signal was measured with a
photomultiplier tube and a lock-in amplifier.}
\end{figure}

In the first experiment we demonstrate enhancement of the TPA rate
by blocking parts of the spectrum of the exciting pulse. To
achieve that, we placed an adjustable slit at the shaper's Fourier
plane, and used it to block spectral bands of the pulse
symmetrically around $\omega_{fg}/2$ (Fig. 2a). The SLM was used
here only for dispersion cancellation in order to produce
transform-limited pulses at the Rubidium gas cell. We measured
both the fluorescence signal and the average power transmitted
through the slit as a function of the "cutoff wavelengths" (i.e.
the shortest and longest wavelengths that passed the slit; see
Fig. 2a).\\

The experimental results are presented in Fig. 2b together with
the theoretical curve calculated by Eq. (\ref{eq3}). When the
cutoff frequencies approached the frequencies of the resonant
transitions, we observed a steep enhancement of the fluorescence
signal, reaching a factor of 2, while the power of the pulse at
that point was reduced by $71\%$. The maximum signal was achieved
when the cutoff frequency was shifted from $\omega_{ig}$ by the
spectral resolution of the system. Closing the slit further
decreased the fluorescence rapidly, which approached zero when the
slit closed completely.\\

\begin{figure}[tbp] \label{fig2}
\begin{center}
\includegraphics[width=8.6cm]{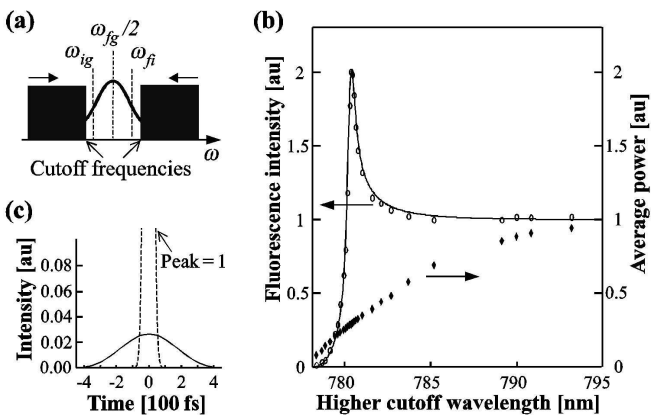}
\end{center}
\caption{Experimental and calculated results for enhancement of
resonant TPA in Rb by selectively blocking parts of the spectrum.
(a) An adjustable slit was used to block spectral bands of the
exciting pulse symmetrically around $\omega_{fg}/2$. (b) The
average power transmitted through the slit (diamonds) together
with the experimental (circles) and calculated (line) normalized
fluorescence intensity as a function of the higher cutoff
wavelength. When the cutoff wavelengths approached the resonant
transitions wavelengths
(to within the spectral resolution of our setup), the average
power was reduced by $71\%$, whereas the TPA rate was doubled. (c)
Calculated temporal intensities of the optimal shaped pulse
(solid) and the initial 50 fs FWHM transform limited pulse
(dashed). The optimal pulse is wider by almost a factor of 8 (390
fs FWHM), and its peak intensity is reduced by a factor of 38.}
\end{figure}

Figure 2c illustrates the calculated temporal intensities of the
optimal shaped pulse and the initial, unshaped transform-limited
pulse, showing both the drastic reduction of the intensity and the
broadening of the pulse. These results demonstrate the
counterintuitive nature of this enhancement. The $71\%$ reduction
in the average power and the 8-fold broadening of the pulse (to
390 fs) result in a reduction of the peak intensity by a factor of
38, yet the TPA rate was increased by $100\%$.

The goal of the next experiment was to achieve maximum enhancement
of the TPA rate by inverting the sign of
$E(\omega)E(\omega_{fg}-\omega)$ about the resonance. Therefore we
used the SLM as a phase filter to apply a phase shift of $\pi/2$
to a 4 nm spectral window, and scanned that window over the
pulse's spectrum. The 4 nm width of the phase window was chosen to
fit the difference between resonant transitions at 780 nm and 776
nm, respectively. Figure 3b shows the measured fluorescence
intensity vs. the spectral position of the phase window, together
with a theoretical curve calculated by Eq. (\ref{eq3}). Maximum
enhancement by a factor of 7 was achieved when the phase window
was centered on $\omega_{fg}/2$, its leading and trailing edges
close to the frequencies of the resonant transitions (Fig. 3a).
Substituting the phase window at that position in Eq. (\ref{eq3})
will show that it performs the desired sign inversion about the
resonance for a spectral region of 8 nm around $\omega_{ig}$, thus
inducing a constructive instead of a destructive interference in
that region. Figure 3c shows the calculated temporal intensities
of the optimal shaped pulse and the initial, unshaped
transform-limited pulse.\

\begin{figure}[tbp] \label{fig3}
\begin{center}
\includegraphics[width=8.6cm]{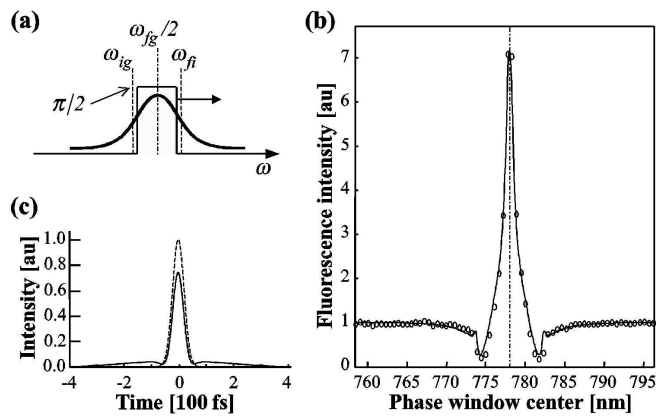}
\end{center}
\caption{Experimental and calculated results for enhancement of
resonant TPA in Rb by scanning a 4 nm $\pi/2$ phase window over
the spectrum of the pulse. (a) The applied phase mask at its
optimal position, centered on $\omega_{fg}/2$, its leading and
trailing edges close to the resonance frequencies. (b)
Experimental (circles) and calculated (line) normalized
fluorescence intensity as function of the spectral position of the
phase window. Maximum enhancement of $600\%$ occurred when the
window was centered on $\omega_{fg}/2$ (778.1 nm, dashed line), as
described in (a), performing the desired sign inversion about
$\omega_{ig}$. (c) Calculated temporal intensities of the optimal
shaped pulse (solid) and the initial transform-limited pulse
(dashed), showing a $26\%$ reduction of the peak intensity due to
the broadening of the pulse.}
\end{figure}

The same enhancement factor (and a similar phase window) was
achieved when we performed an adaptive optimization
\cite{17,18,19,20} of the phase function, using the fluorescence
intensity as a feedback signal.\\

In conclusion, we have shown that resonant multiphoton transitions
can be significantly enhanced by exploiting the general spectral
response of the interaction around resonance. When the interaction
involves an intermediate resonant state, maximizing the peak
intensity by obtaining a transform-limited pulse does not maximize
the transition rate (as is the case with nonresonant
interactions), and pulses with significantly lower intensities can
be more effective. By properly designing the spectral amplitude
and phase of the exciting pulse, we have demonstrated large
enhancements of resonant TPA, in excellent agreement with the
theory. Unlike other experiments in coherent quantum control,
where selectivity between a few processes is the primary goal, we
achieved an enhancement of the absolute rate of a single, simple
nonlinear process. Since this enhancement is based on the general
behavior of any system around resonance, we believe this mechanism
may have played a role in enhancing more complex nonlinear
interactions \cite{18,19}, and could be applied to enhance other
resonant nonlinear processes.


\begin{thebibliography}{}

\bibitem{1} D. J. Tannor and S. A. Rice, J. Chem. Phys. {\bf 83}, 5013 (1985).
\bibitem{2}  M. Shapiro and P. Brumer, J. Chem. Phys. {\bf 84}, 4103 (1986).
\bibitem{3}  D. J. Tannor, R. Kosloff and S. A. Rice, J. Chem. Phys. {\bf 85}, 5805 (1986).
\bibitem{4}  W. S. Warren, H. Rabitz and M. Dahleh, Science {\bf 259}, 1581 (1993).
\bibitem{5}  S. A. Rice, Science {\bf 258}, 412 (1992).
\bibitem{6}  E. D. Potter {\it et al.}, Nature {\bf 355}, 66 (1992).
\bibitem{7}  B. Kohler {\it et al.}, Phys. Rev. Lett. {\bf 74}, 3360 (1995).
\bibitem{8}  A. Hach\'{e} {\it et al.}, Phys. Rev. Lett. {\bf 78}, 306 (1997).
\bibitem{9}  R. N. Zare, Science {\bf 279}, 1875 (1998).
\bibitem{10}  D. C. Clary, Science {\bf 279}, 1879 (1998).
\bibitem{11}  A. M. Weiner {\it et al.}, Science {\bf 247}, 1317 (1990).
\bibitem{12}  D. Meshulach and Y. Silberberg, Nature {\bf 396}, 239 (1998).
\bibitem{13}  D. Meshulach and Y. Silberberg, Phys. Rev. A {\bf 60}, 1287 (1999).
\bibitem{14}  A. M. Weiner and J. P. Heritage, Rev. Phys. Appl. {\bf 22}, 1619 (1987).
\bibitem{15}  A. M. Weiner {\it et al.}, Opt. Lett. {\bf 15}, 326 (1990).
\bibitem{16}  A. M. Weiner, Prog. Quantum Electron. {\bf 19}, 161 (1995).
\bibitem{17}  R. S. Judson and H. Rabitz, Phys. Rev. Lett. {\bf 68}, 1500 (1992).
\bibitem{18}  A. Assion {\it et al.}, Science {\bf 282}, 919 (1998).
\bibitem{19}  R. Bartels {\it et al.}, Nature (to be published).
\bibitem{20}  H. Rabitz {\it et al.}, Science {\bf 288}, 824 (2000).

\end{thebibliography}
\end{document}